# 3-D refractive index tomograms and deformability of individual human red blood cells from cord blood of newborn infants and maternal blood

HyunJoo Park[a,¶], Taegyu Ahn[b,¶], Kyoohyun Kim[a], Sangyun Lee[a], Songyi Kook[b], Dongheon Lee[b], In Bum Suh[c], Sunghun Na[b,*] & YongKeun Park[a,*]

[a]Department of Physics, Korea Advanced Institute of Science and Technology, Daejeon 305-701, Republic of Korea.
[b]Department of Obstetrics and Gynecology, Kangwon National University Hospital, School of Medicine, Kangwon National University, Chuncheon 200-701, Republic of Korea.
[c]Department of Laboratory Medicine, Kangwon National University Hospital, School of Medicine, Kangwon National University, Chuncheon 200-701, Republic of Korea.

**Abstract**
Red blood cells (RBCs) from the cord blood of newborn infants have distinctive functions for fetal and infant development. To systematically investigate the biophysical characteristics of individual cord RBCs in newborn infants, a comparative study was performed of RBCs from cord blood of newborn infants, and of adult RBCs from mothers or non-pregnant women, employing optical holographic micro-tomography. Optical measurements of 3-D refractive index distributions, and of dynamic membrane fluctuations of individual RBCs, enabled retrieval of the morphological, biochemical, and mechanical properties of cord, maternal, and adult RBCs at the individual cell level. The volume and surface area of the cord RBCs were significant larger than those of RBCs from non-pregnant women, and cord RBCs have more flattened shapes than RBCs in adults. In addition, the Hb content in the cord RBCs of newborns was significantly greater. The Hb concentration in cord RBCs was higher than for non-pregnant women or maternal RBCs, but they were within the physiological range of adults. Interestingly, the amplitude of dynamic membrane fluctuations in cord RBCs were comparable to those in non-pregnant women and maternal RBCs, suggesting that the deformability of cord RBCs is similar to that of healthy RBCs in adults.

# 1. Introduction

After birth, a newborn infant starts his or her own circulatory system by disconnection of an umbilical cord that was used for oxygen and nutrient transport from the placenta of the mother. The cord blood left in the umbilical cord of the newborn infant preserves the characteristics of fetal blood at full term, and the characterization of individual cord red blood cells (RBCs) of newborn infants is crucial to understanding maternal-fetal circulation and fetus development, as well as to opening up new possibilities for diagnosing diseases of newborn infants (1).

Previous studies have revealed that cord RBCs of newborn infants are significantly different from those in adult blood. For example, macrocytic RBCs, enlarged RBCs with MCV (mean corpuscular volume) > 110 fL, are predominant in neonatal blood (2), although capillary sizes in newborn infants are similar to those in adults (3). The cytoplasm of cord RBCs is mainly composed of fetal hemoglobin (HbF) rather than adult hemoglobin (HbA); HbF exhibits higher oxygen binding affinity than does HbA (4-6). The life span of cord RBCs (60−80 days) is significantly shorter than normal RBCs (120 days)(7). The aforementioned characteristics of cord RBCs dramatically change within the first 12 weeks after birth (8) by producing RBCs with HbA instead of HbF.

Biochemical characterization of cord RBCs and HbF have been extensively studied, but our understanding of cord RBCs in newborn infants remains incomplete. In particular, the morphological and mechanical properties of cord RBCs in newborn infants, which are closely related to the deformability of cord RBCs and cord-blood circulation, have not been fully addressed largely owing to limitations in measurement techniques. For example, different techniques have led to different interpretations of the deformability of cord RBCs. Previous work using light scattering or hemolysis techniques had reported that cord RBCs are more deformable than adult RBCs (9, 10), whereas electron spin resonance and filtration studies had concluded that there are no significant differences in cell deformability between cord RBCs and adult RBCs (11-13).

Quantitative phase imaging (QPI), however, circumvents the limitations of previous measurement techniques and offers unique advantages to measure precisely the morphological and mechanical properties of cord RBCs. The QPI techniques allow quantitative, non-invasive measurement of optical phase-delay maps induced by transparent samples (e.g., biological cells and tissues) (14, 15). These QPI techniques have been utilized for the study of the pathophysiology of RBCs, and provide unique advantages (16-19). With 3-D QPI techniques, measurement of 3-D refractive index (RI) distributions of individual RBCs can be carried out, which provide morphological (cell volume, surface area and sphericity) and biochemical (Hb content and concentration) information about individual RBCs. This is because RBC cytoplasm is mainly an Hb solution, and the RI of an Hb solution is linearly proportional to its concentration (20). In addition, dynamic fluctuation in RBC membranes can be precise measured using QPI techniques, which provides information about the biomechanical properties of the membrane cortex and the cytoplasm (21-25). Recently, QPI techniques have been utilized for studying the pathophysiology of RBCs including osmotic changes (26), malaria infection (23, 27, 28), sickle cell disease (19, 29), and ATP-dependent morphological remodeling (21, 30).

Here, we report optical measurements of the morphological, biochemical and mechanical properties of cord RBCs in newborn infants. Employing optical holographic micro-tomography, 3-D RI tomograms and dynamic membrane fluctuation of individual RBCs are non-invasively, and quantitatively, measured at the individual-cell level. From the measured RI tomograms, morphological (cell volume, surface area and sphericity) and biochemical (Hb content and concentration) parameters were retrieved. For a comparative study, optical measurements were performed for RBCs collected from the blood of non-pregnant women, from the cord blood of full-term newborn infants (within 5 minutes after delivery), and from maternal blood. The 3-D RI tomographic maps, and the obtained biophysical parameters, clearly demonstrated the distinctive morphologies of cord RBC: large volumes and more flattened discocyte shapes, while the maternal RBCs showed elliptical shapes by loss of the dimple. In addition, the measured dynamic membrane fluctuations showed that cord RBCs have cellular deformability comparable to RBCs of maternal or non-pregnant women.

## 2. Materials and Methods
### *2.1 Ethics Statement*
Human blood studies were conducted according to the principles of the Declaration of Helsinki and were approved by the responsible ethics committee of Kangwon National University Hospital (IRB project number: 2012-0128, Chuncheon, Republic of Korea) before start. Human blood were collected from both non-pregnant women and healthy term pregnant women with 20 years of age or older after obtaining the written informed consent form which explains the blood can be used for academic research purposes. Blood of non-pregnant women was obtained via regular medical checkups executed at the Health Promotion Center of hospital after approval in accordance with the procedures of IRB for the remaining blood. Maternal blood and cord blood from newborn infant were also collected after receiving both the written consent directly from both the mother on behalf of herself and her baby and the verbal consent from the caretakers or the guardians of the minors/children enrolled in the study. This consent form was also approved by the examined the research plan and consent of the committee on the IRB. The collecting method for cord blood has minimum risk to the babies or their mothers because cord blood was collected from the umbilical cord in the placenta after the umbilical cord of the baby was detached from the placenta.

### *2.2 Maternal and cord blood collection*
Blood samples were collected for healthy term pregnant women in Kangwon National University Hospital. Cord blood collections were performed by an obstetrician, who attending the delivery. The umbilical vein of a newborn infant was punctured within 5 minutes after delivery, and 2 mL of cord blood was collected in anticoagulant agent K2 EDTA (ethylenediaminetetraacetic acid) vacutainer (BD, NJ,USA). The maternal blood was sampled on the day and non-pregnant women blood was obtained from remaining blood after examination of health promotion center at Kangwon National University Hospital. All subjects were healthy without any complications. Total 3 individual of non-pregnant women as control, and 5 pairs of newborn infants and their mothers were examined. For the optical measurement, bloods were further diluted 300 times in Dulbecco's PBS buffer (Gibco®, New York, U.S.A.).

### *2.3 Holographic Optical micro-tomography*
The optical setup of holographic optical micro-tomography is described more detail in previous study (31). Briefly, the diode-pumped solid state laser ($\lambda$ = 532 nm, 50 mW, Cobolt, Solna, Sweden) was used as an illumination source. Using rotating a two-axis galvanometer mirror (GVS012/M, Thorlabs, USA), the angle of incident beam was varied. For 3-D RI tomography, the optical fields at various incident illumination angles were measured by employing the common-path laser-interferometric microscopy. The sample, diluted blood sandwiched between two cover glasses with 25×50 mm (C025501, MATSUNAMI GLASS Ind., LTD., JAPAN), is placed between the condenser lens (UPLSAPO 60×, numerical aperture (N.A.) = 0.9, Olympus, Japan) and objective lens (UPLSAPO 60×, N.A. = 1.42, Olympus, Japan). The second galvanometer mirror reflected the beam from a sample to have same optical path regardless of incident illumination angle. After the second galvanometer mirror, a diffraction grating (70 grooves mm$^{-1}$, #46-067, Edmund Optics Inc., NJ, U.S.A.) spatially split the scattering beams and then, spatially filtered 0$^{th}$ order beam as a reference was interfered with the 1$^{st}$ order beam as a sample beam. Then, interferograms were recorded on high-speed sCMOS camera (Neo sCMOS, ANDOR Inc., Northern Ireland, UK) while the incident beam was scanning spirally with 300 different angles. The total magnification was 240 by an additional 4-*f* system. From measured optical fields, 3-D RI distribution of sample was reconstructed using optical diffraction tomography algorithm, found elsewhere (32).

### *2.4 Analysis of the red cell parameters*
The six red cell parameters are comprised of morphological (cell volume, surface area and sphericity), chemical (Hb content and Hb concentration), and mechanical (membrane fluctuation) parameters. To measure the morphological parameters, we used the reconstructed 3-D RI maps by the diffraction optical tomography algorithm from measured multiple optical phase maps corresponding to various illumination angles on the sample. The whole volume of a RBC was calculated by integrating all voxels inside individual a RBCs. The space corresponding to cytoplasm of a RBC was selected by RI with a higher value than threshold. The threshold was defined by 50% of RI difference between the maximum RI of the cell $n_{cell\_max}$ and surrounding medium $n_m$ for determine the cell boundary, i.e. $n_{thresh} = n_m + 0.5 \cdot (n_{cell\_max} - n_m)$. Then, the total number of voxels was multiplied by the magnification of the optical system to translate in a length scale. Next, for surface area measurements, the isosurfaces of individual RBCs were reconstructed from volume data of 3-D RI maps using MATLAB. Surface area of isosurface was measured through the sum of the areas of all the patch faces, which were broken down into small triangular pieces. In addition, the sphericity $SI$, a dimensionless quantity ranging from 0 to 1, was obtained by calculating $SI = \pi^{1/3}(6V)^{2/3}/A$ where the $V$ is the volume and $A$ is the surface area (33, 34).

For the measurement of Hb content, the measured 2-D phase at the normal angle was used. The Hb content of a RBC was obtained from the integrating 2-D optical phase over entire cell area with RI increment of proteins and given as

$$\text{Hb content} = \frac{\lambda}{2\pi\alpha}\sum\Delta\phi(x,y), \qquad (1)$$

where $\lambda$ is the wavelength of laser light (532 nm), $\alpha$ is RI increment (0.2 mL/g) (35, 36) and $\Delta\phi(x,y)$ is 2-D optical phase. In addition, the Hb concentration in a RBC was obtained from the Hb content divided by the cellular volume.

The dynamic membrane fluctuations in RBCs can be quantitatively and precisely measured using the cDOT. Consecutive dynamic full-field optical phase images of a RBC $\Delta\phi(x, y, t)$ can be measured with normal laser illumination, from which dynamic height maps of the RBC can be calculated as, $h(x, y, t) = [\lambda/(2\pi \cdot \Delta n)]\Delta\phi(x, y, t)$, where $\lambda$ is the wavelength of the illumination laser and $\Delta n = \langle n(x, y, z)\rangle - n_m$ is a difference between the mean RI of RBC cytoplasm $\langle n(x, y, z)\rangle$ and surrounding buffer medium $n_m$.

To measure the mechanical parameter, we calculated the dynamic membrane fluctuation from the successively measured the instantaneous height map $h(x,y;t)$, given as:

$$h(x, y;t) = \frac{\lambda}{2\pi\Delta n} \cdot \Delta\phi(x, y;t) \qquad (2)$$

The values for the membrane fluctuation were calculated by averaging the root-mean-square of height displacement over the cell area and given as:

$$\Delta h_{rms}(x, y) = \left\langle \left[h(x, y;t) - h_m(x, y)\right]^2 \right\rangle^{1/2}, \qquad (3)$$

where $h_m$ is the time averaged height at the cell surface.

### 3. Results
#### *3.1 3-D RI tomograms of individual cord and maternal RBCs*
To investigate the 3-D morphological details of individual RBCs, we employed common-path diffraction optical tomography (cDOT). The cDOT is an optical system for holographic micro-tomography, which has the capability of measuring the 3-D RI distribution of a sample with high precision (31). The cDOT measures multiple 2-D optical fields of a sample from different illumination angles, from which the 3D RI tomogram of the sample $n(x, y, z)$ is reconstructed using a DOT algorithm (see *Material and Methods*). Using the cDOT, the 3-D RI distributions of individual RBCs were measured. Samples were collected from three non-pregnant women (121 RBCs); and from five full-term newborn infants (215 RBCs) and their mothers (181 RBCs), after delivery. Then the RBCs were subjected to quantitative, non-invasive measurement.

The 3-D RI maps of characteristic RBCs from each group are shown in Figs.1A–F. The cord RBCs from newborn infants are enlarged compared to the RBCs from non-pregnant women. The RBCs from corresponding mothers, maternal RBCs, are smaller than the cord RBCs, and even smaller than the RBCs from non-pregnant women. Interestingly, the inner dimple area was absent in maternal RBCs while the cord RBCs were biconcave discocytes. The characteristic doughnut shapes and loss of biconcave areas can easily be seen in the rendered isosurfaces of the 3-D RI maps (Figs. 1D–F).

*3.2 Quantitative morphological parameters of individual cord and maternal RBCs*

For quantitative analysis, we calculated the morphological parameters from the measured 3-D RI tomograms of the cord and maternal RBCs. These morphological parameters were cellular volume, surface area, and sphericity (see *Material and Methods*). The cord RBCs of newborn infants exhibited cells that were significantly larger than RBCs from non-pregnant women, in terms of both volume and surface area (Figs. 2A−B). For cord RBCs, the mean values of the cell volume and surface area were $99.5 \pm 16.8$ fL and $181.8 \pm 21.9$ μm$^2$; whereas those of the RBCs from mothers and non-pregnant women were $89.6 \pm 8.0$ fL and $139.2 \pm 17.4$ μm$^2$, and $87.2 \pm 13.0$ fL and $148.9 \pm 16.5$ μm$^2$, respectively. The volumes of RBCs from non-pregnant women measured using cDOT, were consistent with physiological range and the mean cell volumes (MCV), which were independently measured using automated blood-cell counters, based on the complete blood count (CBC). These are indicated by gray lines in Fig. 1. Note that the surface area cannot be obtained from CBC measurements. The mean corpuscular volume with RDW (red cell distribution width) measured by the CBC were $87.3 \pm 15.3$, $99.2 \pm 14.4$, and $89.6 \pm 12.2$ fL for non-pregnant adult RBCs, cord RBCs and maternal RBCs, respectively. The volume and surface area of maternal RBCs were comparable to those of the RBCs from non-pregnant women (Fig. 2C). The mean values of the cell volume and surface area were $89.4 \pm 16.4$ fL and $139.2 \pm 17.4$ μm$^2$. The increased volume in cord RBCs is consistent with previous CBC measurements (2).

In order to analyze the degree of biconcave shape in the cord and maternal RBCs quantitatively, we calculated the sphericity from the measured cell volume and surface area. The sphericity is a dimensionless measure of how spherical an object is. The sphericity of a perfect sphere is '1' and that of a flat surface is '0'. The mean sphericity value of the RBCs from non-pregnant women, exhibiting a characteristic biconcave shape, was $0.64 \pm 0.06$. The sphericity of cord RBCs and maternal RBCs were $0.57 \pm 0.07$, and $0.70 \pm 0.09$, respectively. This indicates that in comparison with healthy RBCs from non-pregnant women, the enlarged cord RBCs exhibit more flattened shapes, and the maternal RBCs are more spherical.

*3.3 Cellular Hb content and concentration in individual cord and maternal RBCs*

To quantify the biochemical characteristics of RBC cytoplasm, Hb content and concentration in the cord and maternal RBCs were quantified from the measured 3-D RI maps. The Hb content of individual RBCs was retrieved from the measured 2-D optical field of cells using the cDOT. Then, the Hb concentration of individual RBCs could be calculated from the Hb content and cell volume of individual RBCs (see *Materials and Methods*).

The mean values of Hb content were $29.5 \pm 4.7$, $35.9 \pm 6.9$, and $30.3 \pm 5.2$ pg for non-pregnant adult RBCs, cord RBCs, and maternal RBCs, respectively (Fig. 3A). The Hb content in cord RBCs was 22% greater than that of RBCs from non-pregnant women, while the Hb content of the maternal RBCs was comparable to that of the RBCs from non-pregnant women. This result, obtained with the cDOT, is also consistent with the CBC measurements: mean corpuscular hemoglobin values were $29.6 \pm 1.2$, $34.7 \pm 0.6$, and $30.7 \pm 3.9$ pg for the non-pregnant adult RBCs, cord RBCs, and maternal RBCs, respectively (gray dotted lines in Fig. 3A).

The mean values of Hb concentration were $33.8 \pm 2.6$, $36.3 \pm 4.7$, and $34.2 \pm 3.3$ g/dL for non-pregnant adult RBCs, cord RBCs, and maternal RBCs, respectively (Fig. 3B). The mean value of Hb concentration of the cord RBCs was measurably greater than that of RBCs from non-pregnant women, and of maternal RBCs.

As shown in Fig. 3C, the correlations between Hb content and cellular volume were positive in all three RBC groups. The linear slope in the correlation map indicated the Hb concentration of each RBC group, and the values of slope were 0.334 ± 0.005, 0.360 ± 0.06, and 0.337 ± 0.004 pg/fL for the RBCs from non-pregnant women, cord RBCs, and maternal RBCs, respectively. These results are also in good agreement with the CBC measurements: mean corpuscular Hb concentration was found to be 33.9, 34.6, and 33.34 g/dL for the RBCs from non-pregnant women, cord RBCs and maternal RBCs, respectively (gray dotted lines in Fig. 3B). Although the Hb concentration in cord RBCs were slightly greater, all the mean values of Hb concentration were within the reference range of a healthy adult (33–36 g/dL).

*3.4 Cellular elasticity of individual cord and maternal RBCs*
To investigate the mechanical properties or deformability of individual cord and maternal RBCs, the dynamic membrane fluctuations of the RBCs were measured. Due to the soft and elastic properties of membrane cortex structures, RBCs exhibit dynamic membrane fluctuations driven by thermal or metabolic energy (21, 37-41). The dynamic membrane fluctuation manifests the deformability of RBC membranes, which is strongly correlated with the structures of the lipid membrane and spectrin network, and alternations caused by various pathophysiological conditions (17, 19, 32).

The dynamic membrane fluctuations in RBCs were quantitatively and precisely measured using the cDOT. Consecutive dynamic full-field optical phase images of individual RBCs were measured with normal laser illumination, from which the mean and dynamic height maps of RBCs were calculated (see *Materials and Methods*). The representative mean and dynamic height maps of individual RBCs in each group are presented in Figs. 4A-C. Consistent with the results of the 3-D rendered isosurfaces (Figs. 1D-F), the mean-cell-shape results demonstrate that the maternal RBCs exhibit spherical shapes without center-dimpled regions, whereas the characteristic donut shapes occur in the cord RBCs as well as in the RBCs from non-pregnant women. The instantaneous displacement maps of the dynamic membrane fluctuations are shown in Fig. 4D-F.

In order to quantify the deformability of individual RBCs, the membrane fluctuations were calculated as the spatially averaged root-mean-squared (*RMS*) height displacement (Fig. 4G). The representative membrane fluctuations of an RBC from a non-pregnant adult, cord blood and maternal blood are presented in Figs. 4D-F, respectively. The *RMS* height displacements of the cord RBC are homogenous over the cell area, and compatible with those of non-pregnant adult RBC and maternal RBC. The mean values of membrane fluctuation were 52.7 ± 5.8, 50.9 ± 5.9, and 52.7 ± 7.1 nm for RBCs from non-pregnant women, the cord, and maternal RBCs, respectively. There were no statistical differences in membrane fluctuation among the three groups of RBCs, indicating that cellular deformability of the cord RBCs and the other RBCs are not significantly altered, despite considerable differences in cell volume and surface area.

**4 . Discussion**
We presented the measurements of morphological, biochemical and mechanical characteristics of individual cord RBCs and maternal RBCs. Quantitative, non-invasive measurements of RBCs using the cDOT, 3-D RI maps and dynamic membrane fluctuations were made. From these, the following important red cell parameters were retrieved systematically: cell volume, surface area, sphericity, Hb content, Hb concentration, and membrane fluctuation. The measured values of cell volume, Hb content, and Hb concentration were consistent with independent CBC measurements, and also consistent with previous reports (7). To date, this is the first reported experimental measurement of surface area, sphericity, and membrane fluctuation of cord and matching maternal RBCs.

We should note that the cDOT measurements provide more detailed information about individual RBCs from cord blood, compared to existing CBC blood tests. For example, the cDOT provides visualization of structural details of the RBC (e.g., 3D shape, surface area, and sphericity), whereas the CBC measurement only provides cell sizes from impedance measurements. The cDOT also provides chemical information about individual RBCs, whereas the CBC test measures ensemble averaged information. Furthermore, deformability measurements are not available using CBC blood tests.

Our results using the cDOT clearly show that the cord RBCs of full-term newborn infants exhibit significant different morphology from the RBCs of non-pregnant women. The volume and surface area of the cord RBCs were 14% and 30% larger, respectively, than those of RBCs from non-pregnant women. The sphericity of the cord RBCs was 11% less than that of the RBCs from non-pregnant women, indicating that cord RBCs have more flattened shapes. The Hb content in the cord RBCs of newborns was significantly greater: the Hb content of the cord RBCs was 22% and 18% greater than that of non-pregnant adult or maternal RBCs. In addition, the Hb concentration in cord RBCs was higher than for non-pregnant women or maternal RBCs, but they were within the physiological range of adults. Interestingly, the amplitude of dynamic membrane fluctuations in cord RBCs were comparable to those in non-pregnant women and maternal RBCs, suggesting that the deformability of cord RBCs is similar to that of healthy RBCs in adults.

It is speculated that these differences in cord RBCs might have resulted from evolution in order to meet the demand for high oxygen consumption by the fetus. Enlarged cell volume with higher Hb concentration in the cord RBCs could carry oxygen to fetal tissues more efficiently. Because the high oxygen-binding affinity of HbF in cord RBCs facilitates the transport of oxygen between two different circulatory systems via an umbilical cord, unique morphological and biochemical properties of cord blood enhance oxygen transport from placenta to fetus.

The question then arises whether this remodeling of cord RBCs is beneficial or detrimental to fetal blood circulation. Although this question is not directly accessible to current experimental study, our measurements of membrane fluctuations suggest that cord RBCs may not be significantly different from other healthy RBCs in their ability to pass through small capillaries and restore their original shapes. This is based on their remarkably soft and elastic properties. Despite the enlarged cell volumes in cord RBCs, their decreased sphericity indicates more discocytic shapes than exhibited by healthy RBCs in adults, suggesting that cord RBCs may still have good ability in passing through narrow passages.

The measured dynamic membrane fluctuations of the cord RBCs, the maternal RBCs, and the non-pregnant adult RBCs are not significantly different from one another. This result is consistent with a previous study using osmotic frangibility which reports indistinguishable cellular deformability between full-term newborn infants and adults RBCs (9). However, considering that other macrocytic RBCs produced due to various clinical situations, including anemia and chronic alcoholism (42, 43), exhibited more deformable RBCs, it is intriguing that the cord RBCs with enlarged cell volumes exhibit dynamic membrane fluctuations comparable to healthy RBCs. In addition, the maternal RBCs exhibit loss of characteristic dimple shapes and became sphere-like. However, the loss of dimple shapes found in ATP-depleted RBCs accompanied decreases in membrane fluctuations, which was not the case in the measured maternal RBCs. Taken together, our results imply complex remodeling in membrane cortex structures in cord RBCs and maternal RBCs.

We presented the optical measurements of morphological, biochemical and mechanical properties of individual cord RBCs and performed comprehensive comparative analysis of maternal RBCs and non-pregnant healthy RBCs. The present method will open possibilities for diagnosis of diseases of newborn infants and their mothers, as well as for the study of pathophysiology of cord RBCs and their implications in fetal circulation. From the technical point of view, the use of the quantitative phase imaging unit can convert an existing microscope into a quantitative phase microscopy (44, 45), and will further expand the applicability of the present technique.

**ACKNOWLEDGEMENTS**
This work was supported by from 2014 Kangwon National University Hospital Grant, KAIST-Khalifar University Project, APCTP, and National Research Foundation (NRF) of Korea (2012R1A1A1009082, 2012-M3C1A1-048860, 2013R1A1A3011886, 2013M3C1A3063046, 2013K1A3A1A09076135, 2014M3C1A3052537, 2014K1A3A1A09063027).

**AUTHOR CONTRIBUTIONS**

Y.P and S.N. developed the experimental idea. H.P. performed the optical experiments and analyzed the data. S.N. prepared the sample of maternal and cord blood. S.N. and Y.P. conceived and supervised the study. All authors discussed the experimental results and wrote the manuscript.

**COMPETING FINALCIAL INTERESTS**
The authors declare no competing financial interests.

**Figures with legends**

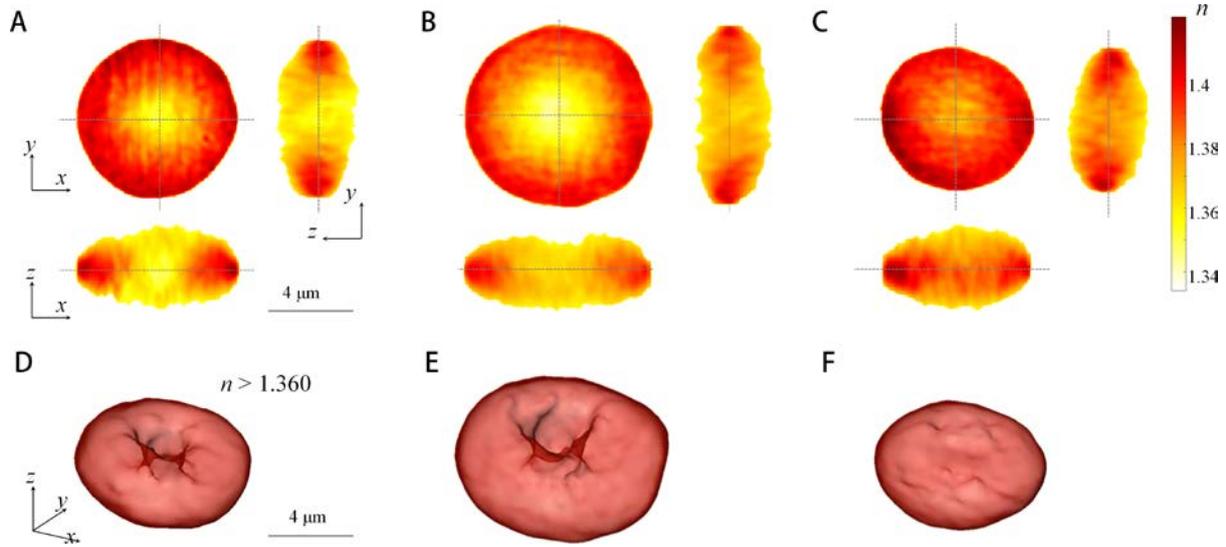

**Fig. 1** A-C, Reconstructed 3-D RI tomograms of RBCs from a healthy non-pregnant women (A), a RBC from cord blood (B), and from a mother (C), respectively. The cross-sectional slices of the RI tomograms are shown in the *x-y* (top left panel), *y-z* (right panel), and *x-z* (bottom panel) planes. D–F, 3-D rendered isosurface images of the RBCs in A–C.

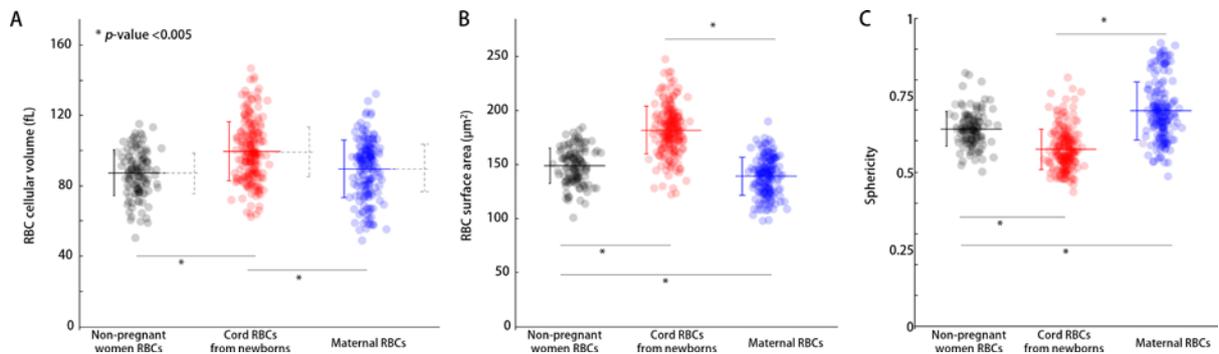

**Fig. 2** Red cell indices for non-pregnant women RBCs ($N = 121$, black circles), cord RBCs from full-term newborn infants ($N = 215$, red circles), and their maternal RBCs ($N = 181$, blue circles): RBCs cellular volume (A), RBCs surface area (B) and sphericity index (C). Each symbol represents an individual RBC measurement and the horizontal solid line is the mean value, and the vertical lines, STD error bars. Gray dot lines in (A) correspond to averaged MCV from the relevant CBC blood test, with vertical lines of RDW (red cell distribution width). The symbol * indicates a *p*-value <0.005.

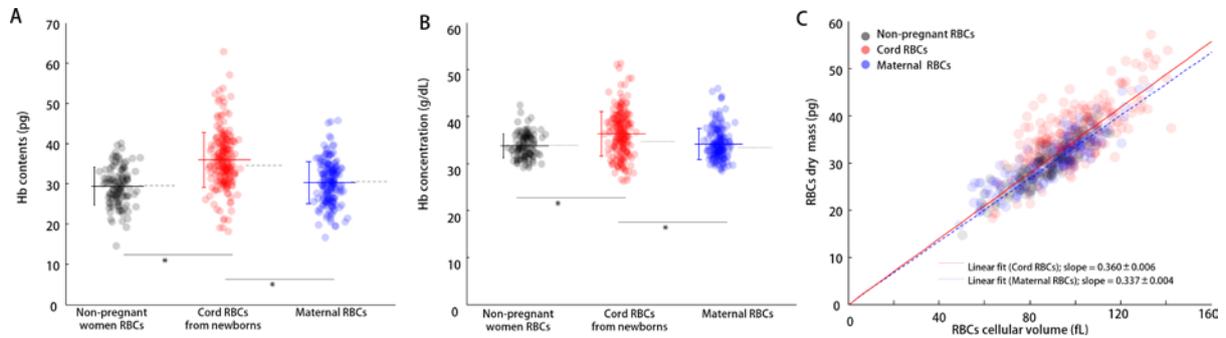

**Fig. 3** RBC dry mass (A), and Hb concentration (B), of non-pregnant women RBCs ($N$ = 121, black circles), fetal cord RBCs ($N$ = 215, red circles) and maternal RBCs ($N$ = 181, blue circles). Each symbol represents an individual RBC measurement and the horizontal line is the mean value, vertical lines indicate STD error bars. Horizontal gray dot lines in (A) and (B) correspond to averaged MCH and MCHC from the CBC blood test. C, Correlation map of RBC cellular volume and RBC dry mass with fitted linear slopes for cord RBCs (red solid line) and maternal RBCs (blue dashed line). The symbol * indicates a $p$-value <0.005.

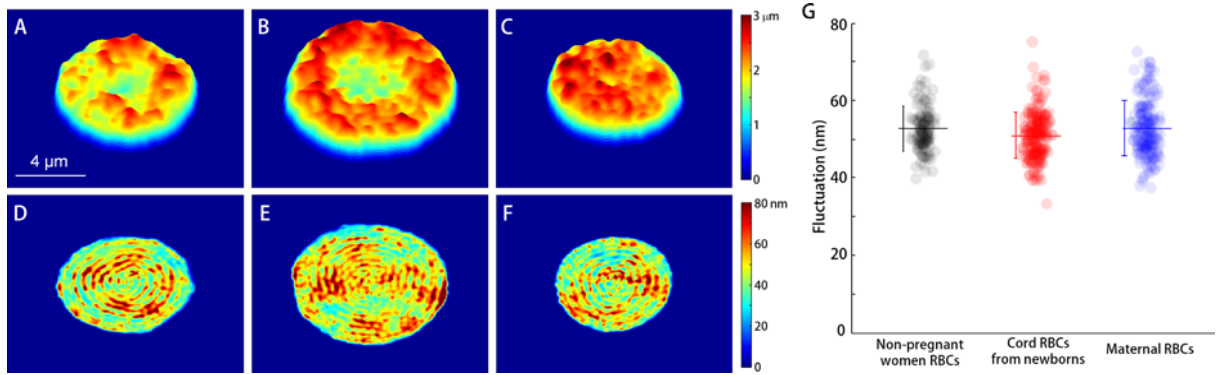

**Fig. 4** (A-C) Representative 2-D topographic images of RBCs from the non-pregnant women, the cord blood of a newborn infant and a mother, respectively. (D-F) Their corresponding dynamic membrane fluctuations. The color bar scales are in μm (top row) and nm (bottom), respectively. (G) The averaged membrane fluctuations of individual RBCs in each group: non-pregnant women (black circles), cord blood of full-term newborn infants (red circles) and their mothers (blue circles). The horizontal solid line is the mean value; the vertical lines are STD error bars.